\documentclass[conference]{IEEEtran}
\usepackage{cite}

\usepackage{enumitem}
\IEEEoverridecommandlockouts

\usepackage{amsmath,amssymb,amsfonts}
\usepackage{algorithmic}
\usepackage{graphicx}
\usepackage{textcomp}
\usepackage{xcolor}
\def\BibTeX{{\rm B\kern-.05em{\sc i\kern-.025em b}\kern-.08em
    T\kern-.1667em\lower.7ex\hbox{E}\kern-.125emX}}

\usepackage[utf8]{inputenc} 
\usepackage[T1]{fontenc}    
\usepackage{hyperref}       
\usepackage{url}            
\usepackage{booktabs}       
\usepackage{amsfonts}       
\usepackage{nicefrac}       
\usepackage{microtype}      
\usepackage{xcolor}         

\usepackage{algorithm}
\usepackage{algorithmic}
\usepackage{amsmath}
\usepackage[skip=0pt]{caption}

\usepackage{tabularx}
\usepackage{subcaption}
\usepackage{siunitx}
\usepackage{latexsym}
\usepackage{amssymb}
\usepackage{graphicx}
\usepackage{caption}
\usepackage{siunitx}
\usepackage{float}
\usepackage{lipsum} 

\usepackage{adjustbox}

\usepackage{booktabs}
\usepackage{array}    
\usepackage{caption}  
\begin{document}

\title{VectorSearch: Enhancing Document Retrieval with Semantic Embeddings and Optimized Search\\
}

\author{
\IEEEauthorblockN{Solmaz Seyed Monir}
\IEEEauthorblockA{\textit{University of Washington} \\
solmazsm@uw.edu}
\and
\IEEEauthorblockN{Irene Lau}
\IEEEauthorblockA{\textit{University of Washington} \\
ixjl@uw.edu}
\and
\IEEEauthorblockN{Shubing Yang}
\IEEEauthorblockA{\textit{University of Washington} \\
sueyoung@uw.edu}
\and
\IEEEauthorblockN{Dongfang Zhao}
\IEEEauthorblockA{\textit{University of Washington} \\
dzhao@uw.edu}
}



\maketitle

\begin{abstract}
Traditional retrieval methods have been essential for assessing document similarity but struggle with capturing semantic nuances. Despite advancements in latent semantic analysis (LSA) and deep learning, achieving comprehensive semantic understanding and accurate retrieval remains challenging due to high dimensionality and semantic gaps. The above challenges call for new techniques to effectively reduce the dimensions and close the semantic gaps. To this end, we propose VectorSearch, which leverages advanced algorithms, embeddings, and indexing techniques for refined retrieval. By utilizing innovative multi-vector search operations and encoding searches with advanced language models, our approach significantly improves retrieval accuracy. Experiments on real-world datasets show that VectorSearch outperforms baseline metrics, demonstrating its efficacy for large-scale retrieval tasks.
\end{abstract}


\section{Introduction}
\subsection{Background and Motivation}
With the exponential growth of digital text data, efficient methods for searching and retrieving relevant information have become increasingly important. Traditional keyword-based search techniques often struggle to capture the semantic meaning of text, leading to suboptimal search results \cite{bibi2023reusable}. The increasing volume of unstructured data, spanning diverse media types like images, videos, textual content, as well as various records such as medical data and real estate information, is largely fueled by its extensive use across multiple domains. This surge can be attributed to the widespread adoption of smartphones, smart devices, and various social networking platforms. According to IDC forecasts, by 2025 \cite{timothy2023toward}, unstructured data is poised to dominate the data landscape, constituting a staggering 80\% \cite{king80percent, wang2021milvus} of total data volume. This exponential growth, concurrent with the rapid advancements in machine learning, underscores the critical necessity for robust methodologies aimed at converting this unstructured data into feature vectors. Vector embedding, a prevalent technique harnessed in recommender systems for transforming raw data into structured feature vectors, has gained substantial traction in recent years. However, existing paradigms in vector data management predominantly center on vector similarity search, encountering significant challenges in meeting evolving demands due to their inherent limitations, including constrained support for multi-vector queries and suboptimal performance, particularly in handling large-scale and dynamically evolving vector datasets \cite{wang2021milvus}.
The connection between efficient information retrieval and vector databases is based on the ability of vector embeddings to capture complex semantic relationships within data. This capability is essential for developing advanced IR systems that can provide more accurate and contextually relevant results. However, existing research on vector data management, as documented in prior studies \cite{lu2020r2lsh, lu2020vhp, lv2017intelligent}, predominantly centers around enhancing vector similarity search capabilities. Nonetheless, these approaches confront difficulties in addressing evolving needs due to their limited functionalities, such as inadequate support for handling multi-vector queries, and subpar performance in dealing with large-scale and dynamically changing vector datasets \cite{wang2021milvus}.

Current systems, such as Milvus \cite{wang2021milvus}, offer multi-vector support and are optimized for large-scale vector data management. However, even with Milvus's distributed architecture, handling dynamically changing datasets—especially in environments with high-dimensional data—can introduce performance challenges, due to the reindexing overhead and query latency. Moreover, while Retrieval Augmented Generation (RAG) \cite{lewis2020retrieval} frameworks effectively integrate retrieval mechanisms with generative models, they are not specifically designed for real-time updates and high-dimensional indexing in distributed systems. These frameworks do not fully leverage hybrid indexing techniques for optimal performance in multi-vector search.

In focusing solely on algorithms, we uncover several limitations of vector similarity search algorithms \cite{lu2020r2lsh, lu2020vhp, lv2017intelligent, malkov2018efficient}. Many methodologies and libraries depend heavily on main memory storage and lack the capability to distribute data across multiple machines, thereby hindering scalability \cite{wang2021milvus, elasticsearch}. Additionally, current algorithms are predominantly designed for static datasets and struggle to accommodate dynamic data. This restriction significantly impacts their adaptability to real-world scenarios where data is constantly changing. Moreover, the absence of advanced query processing capabilities in existing solutions further curtails their practical applicability \cite{wang2021milvus}. Sophisticated query processing is essential for handling complex queries involving multiple vectors, which are common in many information retrieval tasks.

Addressing these limitations is crucial for developing more effective and scalable vector data management systems, which can better support the complex and evolving needs of modern information retrieval applications. Advancements in this area will enable more robust and efficient IR systems.
\subsection{Proposed Approach}
Our proposed approach, VectorSearch, represents a novel advancement in the realm of information retrieval. It operates as a hybrid system, combining the strengths of vector embeddings and traditional indexing techniques to overcome various limitations encountered in existing algorithms and systems. By integrating advanced methods such as FAISS for efficient distributed indexing, VectorSearch enables seamless management of large-scale datasets across multiple machines. Additionally, VectorSearch incorporates HNSWlib for further optimization and enhancement of search capabilities, ensuring efficient retrieval of relevant documents even in dynamic and evolving environments. This hybrid format empowers VectorSearch to deliver superior performance, scalability.
Our VectorSearch algorithm is uniquely designed to handle dynamic data with mechanisms for multi-vector query handling, we enable advanced query processing, facilitating sophisticated search operations beyond mere similarity searches. Leveraging embeddings and vector databases for multi-vector search, we encode text data into high-dimensional embeddings $\mathbf{E} = \{ \mathbf{e}_1, \mathbf{e}_2, \ldots, \mathbf{e}_n \}$ and index them in a vector database \cite{gunther2018freddy}, enabling efficient retrieval of relevant text pieces based on user queries. Furthermore, the search operation in VectorSearch involves finding the nearest neighbors of a query embedding vector $\mathbf{q}$. Let $\mathbf{I}(\mathbf{E})$ represent the index structure mapping the embedding vectors to their corresponding positions. The search operation can be denoted as $\text{Search}(\mathbf{q}, \mathbf{I}(\mathbf{E}))$, where $\mathbf{q}$ is the query embedding vector. The result of this operation is a set of embedding vectors representing relevant documents. 

We propose \textbf{VectorSearch}, a hybrid document retrieval framework that integrates advanced language models, multi-vector indexing techniques, and hyperparameter optimization to improve retrieval precision and query time in high-dimensional spaces. Unlike existing solutions, our approach:
\begin{enumerate}
\item We propose innovative \textbf{Multi-Vector} Search algorithms that encode documents into high-dimensional embeddings, significantly optimizing retrieval efficiency. These algorithms leverage advanced techniques in semantic embeddings to represent documents in a high-dimensional vector space (Section III). 
\item We propose an innovative algorithm that optimizes nearest neighbor search using both single- and \textbf{multi-vector strategies}, significantly improving search efficiency (Section III-B).

\item We implement an innovative strategy that systematically tunes index dimension, similarity threshold, and model selection to optimize the retrieval system (Section III-D).
\end{enumerate} 
\section{Related Work}
Previous research on similarity search can be organized into four main categories \cite{wang2021milvus}: tree-based methods \cite{lu2020vhp}, LSH-based techniques \cite{gong2020idec, wang2021milvus, li2020efficient, lu2020r2lsh}, quantization-based approaches \cite{andre2016cache}, and graph-based algorithms \cite{malkov2018efficient}. While previous research primarily focuses on index-centric approaches, VectorSearch distinguishes itself as a comprehensive vector data management system. Beyond mere indexing, VectorSearch integrates query engines, CPU optimization providing a comprehensive solution for efficient and scalable document retrieval. Additionally, it incorporates cache mechanisms \cite{zhao2013hycache, zhao2014hycache+, monir2024efficient} to further enhance performance and response times.

Embedding-driven Retrieval presents formidable hurdles for search engines due to the massive scale of textual data involved. Unlike ranking layers, which typically manage a few hundred items per session, the retrieval layer of a search engine must efficiently process trillions of text documents within its index. This extensive scale poses a dual challenge for search engines, involving both serving and training embeddings tailored for textual content \cite{huang2020embedding}. 

Prior research in the field of vector similarity search has primarily focused on developing algorithms and systems for efficient retrieval of similar vectors. Existing works \cite{lu2020r2lsh, lu2020vhp, lv2017intelligent, malkov2018efficient} along with their associated Faiss. However, these efforts often suffer from several limitations. Firstly, lacking comprehensive systems capable of managing large volumes of vector data efficiently, they struggle to handle datasets that exceed main memory. Additionally, most existing solutions assume static data once ingested, making it challenging to accommodate dynamic updates. Our proposed algorithm addresses these shortcomings by providing a solution that integrates embeddings, vector databases, and mechanisms for multi-vector query handling.

The emergence of various models aimed at enhancing precision and recall in information retrieval tasks. Notably, approaches such as SimCSE, ESimCSE, VaSCL, Prompt-RoBERTa, and CARDS \cite{wang2022improving} have made substantial strides in improving performance across a spectrum of tasks, including STS evaluations. Our research seeks to address these challenges by presenting models like MiniLM-L6-v2 and BERT-base-uncased \cite {brahma2024improving, Pretrain1:online}, which demonstrate competitive precision and recall rates while optimizing query time. NCI \cite{wang2022neural} requires a significantly larger model capacity to extend to web scale. To address this, the VectorSearch utilizes advanced hybrid format indexing techniques which allow for the seamless management of large-scale datasets, providing scalability without the need for excessively large models. NCI needs improvement to serve online queries in real time. VectorSearch enhances the speed of searches by proposed indexing for efficient high-dimensional vector searches. This significantly reduces query times, NCI faces challenges in updating the model-based index when new documents are added. In contrast, VectorSearch is designed to handle dynamic data.      

\begin{figure*}[t]
  \centering
  \includegraphics[width=\textwidth]{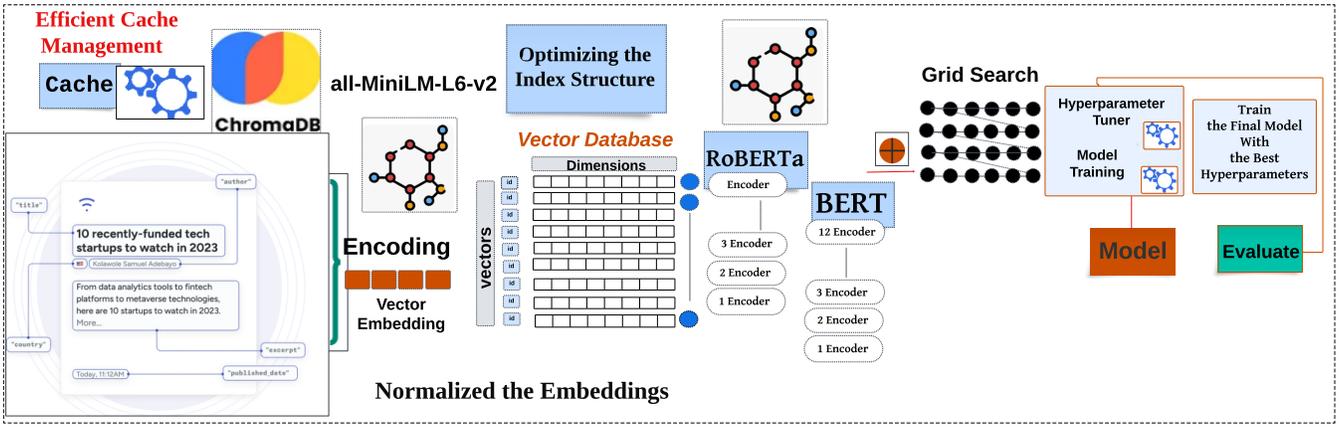}
  \captionsetup{width=\linewidth, font=normalsize}
  \caption{We propose the VectorSearch Framework, utilizing a systematic grid search to fine-tune document retrieval systems by optimizing hyperparameters, index dimensions, and similarity thresholds for enhanced performance.}
  \label{fig:framework}
\end{figure*}
\section{Design and Implemntation}
VectorSearch benefits from a hybrid approach that leverages the strengths of both indexes. HNSWlib's hierarchical \cite{malkov2018efficient} structure enables efficient navigation of high-dimensional semantic embedding spaces. This hierarchical structure organizes embeddings into a navigable graph, enabling fast and accurate similarity search. Let $\mathbf{X} = \{ \mathbf{x}_1, \mathbf{x}_2, \ldots, \mathbf{x}_n \}$ represent the input text corpus, where $\mathbf{x}_i \in \mathbb{R}^d$ is a high-dimensional vector embedding of the $i$-th document. The embeddings are indexed using hybrid format for efficient retrieval. The hierarchical structure organizes embeddings into a graph $G = (V, E)$, where $V$ is the set of vertices corresponding to the document embeddings, and $E$ is the set of edges representing the navigable connections between them. Here, $\text{similarity}(\mathbf{x}_i, \mathbf{x}_j)$ denotes a similarity metric and $\tau$ is a predefined threshold. Given a query embedding $\mathbf{q} \in \mathbb{R}^d$, the search process in HNSWlib can be formulated as finding the $k$-nearest neighbors (k-NN) of $\mathbf{q}$ in the graph $G$. where $q$ is the query embedding vector, $\mathcal{E}$ is the set of embeddings in the graph, and $d(q, e_i)$ is the distance metric used to measure similarity between the query and the embeddings. The hybrid approach which utilizes flat and inverted file indexing methods, provides a broad and efficient search capability, while the index refines this search through its graph-based hierarchical structure, enabling rapid and accurate similarity searches. 

\textbf{Embedding.} We utilized transformer-based models (BERT, RoBERTa) \cite{chatterjee2022bert} to produce embeddings for text data. These embeddings capture semantic information about the text and are high-dimensional vectors.

\textbf{Vector Database.} A vector database, ChromaDB, was utilized \cite{trychromaAInativeOpensource} to index and store the produced embeddings \cite{remis2021using}. This allows for efficient storage and retrieval of high-dimensional vectors \cite{monir2024efficient}.

\textbf{Search Operations.} Given a query, we performed similarity search operations on the indexed embeddings and efficiently retrieves \cite{huang2020effective, wang2022deep} the most similar embeddings \cite{gunther2018freddy} from the vector database based on a cosine similarity metric.\begin{equation}\begin{aligned}G &= (V, E) \quad  \\V &= \{ \mathbf{x}_1, \mathbf{x}_2, \ldots, \mathbf{x}_n \}, \\E &= \{ (\mathbf{x}_i, \mathbf{x}_j) \mid \text{similarity}(\mathbf{x}_i, \mathbf{x}_j) > \tau \}\end{aligned}\end{equation}\begin{equation}\text{Search}(\mathbf{q}, G) = \{ \mathbf{x}_i \in V \mid \allowbreak \mathbf{x}_i \text{ is one of the k-NN of } \mathbf{q} \}\end{equation} \begin{equation}\text{Search}(q, G(\mathcal{E})) = \arg \min_{e_i \in \mathcal{E}} d(q, e_i)\end{equation}
\textbf{Data Preprocessing and Model Initialization.}
The initial step involved preprocessing the dataset obtained from the Newscatcher API and selecting a subset of documents for analysis. To encode the titles of these documents into numerical vectors suitable for vector search, We utilized  the capabilities of the SentenceTransformer, renowned for its ability to produce semantically meaningful embeddings. During model initialization, we implemented a caching mechanism to optimize resource utilization. By designating a cache directory within the local filesystem \cite{zhao2013hycache, zhao2014hycache+}, the library stored precomputed model weights and configurations. This strategy ensured rapid loading of the model and eliminated the need for repetitive downloads, thereby enhancing computational efficiency and minimizing network latency \cite{timothy2023toward}.
\begin{algorithm}[H]
        \caption{\textbf{Proposed VectorSearch Framework}}\label{alg:VectorSearch Framework}
        \begin{algorithmic}[1]
        \REQUIRE Dataset $\mathcal{D}$ containing document titles
        \ENSURE Best hyperparameters $\theta_{\text{best}}$ and evaluation results $Best\_Results$
        \STATE Load the dataset $\mathcal{D}$
        \STATE Process the data and create SentenceTransformer examples
        \STATE \textbf{Encode:} Encode document titles using the SentenceTransformer model: $E = \{e_1, e_2, ..., e_n\}$
        \STATE \textbf{Normalize:} Normalize encoded vectors: $E_{\text{norm}} = \left\{\frac{e_i}{\|e_i\|}\right\}$
        \STATE \textbf{Add to Index:} Add normalized vectors to the FAISS index: $I.add(E_{\text{norm}})$
        \STATE Define $f(\theta) \rightarrow \{precision, recall, query\_time\}$
        \STATE Define hyperparameter grid: $\Theta = \{\theta_1, \theta_2, ..., \theta_n\}$
        \STATE Combinations: $\Theta_{\text{combinations}} = \{\theta_{11}, \theta_{12}, ..., \theta_{mn}\}$
        \FOR{each combination $\theta_{ij}$ in $\Theta_{\text{combinations}}$}
            \STATE Evaluate: $(pre_{ij}, re_{ij}, query\_time_{ij}) = f(\theta_{ij})$
            \STATE Store: $Results_{ij} = \{precision_{ij}, recall_{ij}, query\_time_{ij}\}$
        \ENDFOR
        \STATE  $\theta_{\text{best}} = \arg\max_{\theta_{ij}} precision_{ij}$
        \STATE Retrieve: $Best\_Results = Results\_DF[\theta_{\text{best}}]$
        \end{algorithmic}
        \end{algorithm}
\textbf{Indexing.} We built indexes on vector embeddings to utilize the HNSWlib  \cite{docarrayHnswlibDocument} and FAISS \cite{douze2024faiss} indexes in the VectorSearch framework. These indexes enable fast and accurate retrieval of similar documents by organizing the vector embeddings into efficient data structures, such as navigable graphs HNSWlib and inverted files FAISS, which allow for approximate nearest neighbor (ANN) search \cite{rahman2022evaluating, malkov2018efficient}.

\begin{algorithm}[t]
\caption{\textbf{Proposed Scalable Multi-Vector Search Algorithm}}
\label{alg:multi_vector_search}
\textbf{Input:} FAISS Index $index$, Query Vectors $query\_vectors$, Number of Nearest Neighbors $k$ \\
\textbf{Output:} Set of Similar Vectors $similar\_vectors$
\begin{algorithmic}[1]
    \STATE Initialize FAISS index with dimensionality $dim$
    \STATE $index \leftarrow$ initialize\_faiss\_index($dim$)
    \STATE \textbf{Function} single\_vector\_search($index$, $query\_vector$, $k$):
    \STATE \quad Set $index.nprobe = 10$
    \STATE \quad $results, idx \leftarrow$ index.search($query\_vector$, $k$)
    \STATE \quad \textbf{return} $idx[0]$
    \STATE \textbf{Function} multi\_vector\_search($index$, $query\_vectors$, $k$):
    \STATE \quad $all\_results \leftarrow []$
    \STATE \quad \textbf{for} $query\_vector$ \textbf{in} $query\_vectors$:
    \STATE \quad \quad $results \leftarrow$ single\_vector\_search($index$, $query\_vector$, $k$)
    \STATE \quad \quad $all\_results$.extend($results$)
    \STATE \quad \textbf{return} unique($all\_results$)
    \STATE $similar\_vectors \leftarrow$ multi\_vector\_search($index$, $query\_vectors$, $k$)
    \STATE \textbf{Output} $similar\_vectors$
\end{algorithmic}
\end{algorithm}
\textbf{Query Processing.} We handled user queries by encoding them into vector representations $\mathbf{q} \in \mathbb{R}^d$ and performing similarity search using the indexed vectors $\mathbf{E} = \{\mathbf{e}_1, \mathbf{e}_2, \ldots, \mathbf{e}_n\}$, where each $\mathbf{e}_i \in \mathbb{R}^d$ is an embedding vector. Algorithm \ref{alg:multi_vector_search} leverages the advanced capabilities of multi-vector search, facilitating the efficient retrieval of similar vectors across diverse datasets. By implementing a robust indexing mechanism, our approach establishes a high-performance structure adept at managing high-dimensional vector data \cite{monir2024efficient}.The single-vector search operations within the index are enhanced, extending the methodology to efficiently manage multi-vector queries.
 Specifically, for a multi-vector query $\mathbf{Q} = \{\mathbf{q}_1, \mathbf{q}_2, \ldots, \mathbf{q}_m\}$, our algorithm searches for each query vector $\mathbf{q}_j$ in $\mathbf{Q}$, retrieving the nearest neighbors $\mathbf{N}(\mathbf{q}_j)$ from the indexed vectors. The similarity search operation can be represented as:\begin{equation}\text{Search}(\mathbf{q}, \mathbf{E}) = \arg\min_{\mathbf{e}_i \in \mathbf{E}} d(\mathbf{q},\mathbf{e}_i),\end{equation}
where $d(\mathbf{q}, \mathbf{e}_i)$ is a distance metric. We propose a comprehensive evaluation methodology for assessing the effectiveness of the VectorSearch system ref {alg:VectorSearch Framework}. This methodology involves conducting comprehensive evaluations across diverse hyperparameter configurations. The performance metrics, including mean precision $\bar{P}$ and query time $T_q$, were measured for each combination of hyperparameters $\theta$. We utilized \texttt{ParameterGrid} from the \texttt{scikit-learn} library to systematically explore the hyperparameter space $\Theta$. By iterating over the parameter grid $\Theta = \{\theta_1, \theta_2, \ldots, \theta_k\}$, we identified optimal configurations $\theta^*$ that maximized precision while minimizing query time. The optimization process can be expressed as: \begin{equation}\theta^* = \arg \max_{\theta \in \Theta} \left(\frac{\bar{P}(\theta)}{T_q(\theta)}\right).\end{equation}

\subsection{VectorSearch Design}
We proposed the VectorSearch framework, as shown in Figure \ref{fig:framework}, and Algorithm \ref{fig:framework} as a systematic approach to document retrieval leveraging state-of-the-art techniques. This framework provides a structured methodology for optimizing document retrieval systems, offering insights into the effectiveness of different hyperparameter configurations and facilitating the identification of optimal settings. The Parameter Grid is utilized to define a comprehensive parameter grid, encompassing various combinations of hyperparameters such as pretrained model selection ($\theta_{\text{model}}$), index dimensionality ($\theta_{\text{dimension}}$) and similarity threshold ($\theta_{\text{threshold}}$).

\textbf{Feature Extraction (Embedding).} That utilized a deep learning model, denoted as $\text{Embedding}(\cdot)$, to convert the preprocessed document titles into embeddings. Thus, each document $\mathbf{d}_i$ is transformed into an embedding $\mathbf{e}_i.$
$\mathbf{e}_i = \text{Embedding}(\text{Preprocess}(\mathbf{d}_i))$.

\textbf{Vector Database Creation ($\mathcal{V}$).} 
That constructed a vector database $\mathcal{V}$ using the embeddings of the document titles. The database $\mathcal{V}$ is indexed using the FAISS, facilitating efficient similarity search operations where $(\mathcal{V}): \quad \mathcal{V} = \{ \mathbf{e}_1, \mathbf{e}_2, \ldots, \mathbf{e}_n \}$.
To effectively implement this process, we propose the efficient multi-vector search algorithm \ref{alg:multivector}. Model Training and Evaluation that defined a function rain Evaluate $(\theta)$, where $\theta$ represents the hyperparameters of the VectorSearch framework. This function trains and evaluates the model, returning performance metrics.

\textbf{Hyperparameter Tuning ($\Theta$).} Defined a set of hyperparameters $\Theta = {\theta_1, \theta_2, ..., \theta_M}$, where each $\theta_i$ represents a combination of hyperparameters.

\textbf{Optimization Objective.} Goal is to find the optimal hyperparameters $\theta^*$ and maximizing the precision metric. \begin{equation}\theta^* = \arg\max_{\theta \in \Theta} \text{Precision}(\theta)\end{equation}
\subsection{Scalable Multi-Vector Search}
We peoposed Algorithm \ref{alg:multi_vector_search} scalable multi-vector search approach for retrieving similar vectors efficiently. It utilizes the FAISS index and accepts query vectors along with the desired number of nearest neighbors. The algorithm begins by initializing the FAISS index with a specified dimensionality. Then, it defines two functions: single-vector-search and multi-vector-search. The single-vector-search function conducts a nearest neighbor search for a single query vector, while the multi-vector-search function extends this process to multiple query vectors, aggregating the results. Finally, the algorithm outputs the set of similar vectors retrieved from the multi-vector search. This algorithm complements the VectorSearch design \ref{alg:Vec} by providing a mechanism for efficient multi-vector search operations that document titles are encoded using the SentenceTransformer model, and an index is constructed for efficient similarity search. Systematic evaluation of hyperparameter combinations aids in algorithm fine-tuning. 
The VectorSearch design Algorithm \ref{alg:Vec} involves systematically exploring various hyperparameter combinations to tune the system's settings and optimize the performance of the document retrieval system. For data preprocessing we removed HTML tags, handling missing values, tokenization, removing stopwords, and lemmatization. Document title encoding operates at $\mathcal{O}(n)$ time complexity ($n$: number of titles), while vector normalization incurs $\mathcal{O}(nd)$ time ($d$: embedding dimension). Adding the normalized vectors to the index involves inserting each vector into the index, resulting in a time complexity of $\mathcal{O}(nd)$. Evaluating the performance of different hyperparameter combinations involves iterating over all combinations and evaluating the performance for each combination. This results in a time complexity of $\mathcal{O}(mn)$, where $m$ is the number of hyperparameter combinations and $n$ is the number of documents in the dataset. Overall, the complexity of the VectorSearch Framework algorithm can be summarized as $\mathcal{O}(nd + mn)$, where $n$ is the number of documents, $d$ is the dimensionality of the embeddings, and $m$ is the number of hyperparameter combinations.
\begin{algorithm}[!htbp]
\caption{\textbf{Proposed VectorSearch Algorithm}}\label{alg:Vec}
\begin{algorithmic}[1]
\REQUIRE Queries $Q$, Index $I$, $k$
\ENSURE Results $R$
\FOR{each query $q_i$ in $Q$}
\STATE Encode $q_i$ into vector $\mathbf{q_i}$ using SentenceTransformer model
\STATE Normalize $\mathbf{q_i}$
\STATE Perform nearest neighbor search with $\mathbf{q_i}$ using $I$
\STATE Extract top $k$ results from the search: $\{(d_1, s_1), (d_2, s_2), \dots, (d_k, s_k)\}$
\STATE Retrieve documents corresponding to $d_j$ from the index and assign similarity scores $s_j$ to them
\STATE Append retrieved documents to $R$
\ENDFOR
\RETURN $R$
\end{algorithmic}
\end{algorithm}
\subsubsection{VectorSearch Algorithm and Complexity}
We \textbf{propose a novel approach} \ref{alg:Vec} to document retrieval, leveraging state-of-the-art techniques in natural language processing and information retrieval. Our methodology integrates advanced encoding models such as SentenceTransformer with efficient indexing techniques to enable rapid and accurate retrieval of relevant documents. By systematically exploring various hyperparameter configurations and employing rigorous evaluation methods, we aim to optimize the performance of our document retrieval system. The proposed VectorSearch algorithm outlines the process of conducting a nearest neighbor search within a given index to retrieve relevant documents for a set of queries. Given a set of queries \( Q \), an index \( I \), and a specified number \( k \) for the top results to be retrieved, the algorithm iterates over each query \( q_i \). For each query, the algorithm encodes it into a vector \( \mathbf{q}_i \) using a SentenceTransformer model and normalizes the vector. It then performs a nearest neighbor search with the normalized query vector \( \mathbf{q}_i \) using the index \( I \). The top \( k \) results from the search, each consisting of a document \( d_j \) and its corresponding similarity score \( s_j \), are extracted. The algorithm retrieves the documents corresponding to these top results from the index and assigns the similarity scores \( s_j \) to them. 
\begin{algorithm}[!htbp]
\caption{\textbf{Proposed Training and Evaluation of Vector Search Systems}}
\label{alg:train_eval}
\begin{algorithmic}[1]
\STATE Load data: $pdf \leftarrow \text{Load\_Data}(C)$
\STATE Process data: $pdf\_subset \leftarrow pdf[1:1000]$
\STATE Load model: $model \leftarrow \text{ST}(L)$
\STATE Encode titles: $fe \leftarrow \text{model.encode}(pdf\_subset.t)$
\STATE Create FAISS index: $fi \leftarrow \text{IF}(len(fe[0]))$
\STATE Add vectors: $fi.\text{add}(fe)$
\STATE Create HNSWlib index: $hi \leftarrow \text{hnswlib.Index}(C)$
\STATE Initialize index: $hi.\text{init\_index}(len(fe),C, M)$
\STATE Add items: $hi.\text{add\_items}(fe)$
\STATE Define $\texttt{train\_eval}()$ and $param\_grid$
\STATE $pc \leftarrow \text{PG}(param\_grid)$
\FORALL{$p$ in $pc$}
    \STATE Evaluate: $\texttt{train\_eval}(p)$
\ENDFOR
\end{algorithmic}
\end{algorithm}
\begin{algorithm}[H]
    \caption{\textbf{Proposed Efficient Multi-Vector Search Algorithm}}
    \label{alg:multivector}
    \begin{algorithmic}[1]
        \REQUIRE $\mathcal{Q}$: Query, $\mathcal{D}$: Dataset
        \ENSURE $\mathcal{R}$: Relevant Documents
        \STATE $\mathbf{E} \leftarrow f(\mathcal{D})$ \hfill \textit{\% Encode documents in the dataset into embeddings}
        \STATE $\mathcal{I} \leftarrow \text{create\_index}(\mathbf{E})$ \hfill \textit{\% Initialize a vector database and index the embeddings}
        \STATE $\mathbf{q} \leftarrow f(\mathcal{Q})$ \hfill \textit{\% Encode the query into an embedding}
        \STATE $\mathcal{N} \leftarrow \text{nearest\_neighbors}(\mathbf{q}, \mathcal{I})$ \hfill \textit{\% Perform a nearest neighbor search in the index}
        \STATE $\mathcal{R} \leftarrow \text{retrieve\_top\_k}(\mathcal{N}, k)$ \hfill \textit{\% Retrieve top-$k$ most similar documents}
        \STATE $\text{present\_results}(\mathcal{R})$ \hfill \textit{\% Present relevant documents to the user}
    \end{algorithmic}
\end{algorithm}
 \subsection{Efficient Multi-Vector Search Algorithm}
In the proposed Algorithm \ref {alg:multivector}, the complexity of encoding documents($f(\mathcal{D})$) into embeddings is contingent upon the specific embedding model employed. Let $N$ represent the number of documents in the dataset, and let $d$ denote the dimensionality of the embeddings. The time complexity for encoding all documents is typically $O(Nd)$. 
The complexity of creating Index ($\text{create\_index}(\mathbf{E})$) depends on the indexing algorithm. Let's denote the number of embeddings as $M$. For approximate nearest neighbor methods, the complexity is $\mathcal{O}(M\log(M))$ or $\mathcal{O}(Md)$. Encoding Query ($f(\mathcal{Q})$): Similar to encoding documents, the complexity of encoding the query depends on the specific embedding model and is typically $\mathcal{O}(d)$. 
Performing the nearest neighbor search ($\text{nearest\_neighbors}(\mathbf{q}, \mathcal{I})$) has a complexity that depends on the indexing algorithm used. For hybrid methods, it is typically $\mathcal{O}(\log(M))$ or $\mathcal{O}(\log(M) + k)$, where $k$ is the number of nearest neighbors to retrieve. Retrieving the top-k documents ($\text{retrieve\_top\_k}(\mathcal{N}, k)$) is straightforward when $\mathcal{N}$ contains $k$ nearest neighbors and has a complexity of $\mathcal{O}(k)$.
\subsection{Optimizing Vector Search Systems}
We introduce Algorithm \ref{alg:train_eval} as a comprehensive method for training and evaluating document retrieval systems.
This algorithm focuses on systematically exploring various hyperparameter configurations to optimize the performance of the document retrieval system. Initially,  the algorithm loads a dataset containing document titles and associated metadata. Subsequently, it extracts a representative subset, $pdf_{\text{subset}}$, comprising the initial 1000 documents. Using the encoder function $f_{\text{enc}}$, document titles are transformed into dense vector representations in $\mathbb{R}^d$, resulting in a feature matrix $fe \in \mathbb{R}^{n \times d}$. 
These vectors are integrated into ($fi$) and ($hi$) hybrid index to facilitate efficient nearest neighbor search operations. 
An evaluation function $\texttt{evaluate}(p)$ is defined to quantify system performance metrics and evaluate the system's effectiveness across diverse hyperparameter configurations. 
Algorithm Complexity: Data Loading and Preprocessing: \( O(n) \). Embedding Generation: \( O(nd) \). Index Construction: \( O(nd \log n) \) - \( O(nd) \). Parameter Grid Generation: \( O(pm) \). Model Training and Evaluation: \( O(ks) \).
\section{Experimental Evaluations}
The experiments used a labeled dataset of 1000 news articles. We implemented the algorithm in Python, using libraries for data manipulation, computations, and NLP. SentenceTransformer encoded document titles into embeddings. Indexes facilitated retrieval. Hyperparameter optimization evaluated combinations of dimensions, thresholds, and models using grid search. All our experiments were performed using the same hardware consisting of RTX NVIDIA 3050 GPUs and i5-11400H @ 2.70GHz with 16GB of memory. The details of each experiment are the following. We implemented a caching mechanism to store and reuse precomputed embeddings from the Chroma model, enhancing efficiency by eliminating redundant computations. This mechanism efficiently saved embeddings to disk, minimizing the need for recomputation and optimizing resource management. 
\subsection{Datasets} 
\begin{figure}[ht]
    \centering
    \begin{subfigure}[b]{0.61\linewidth} 
        \includegraphics[width=\linewidth]{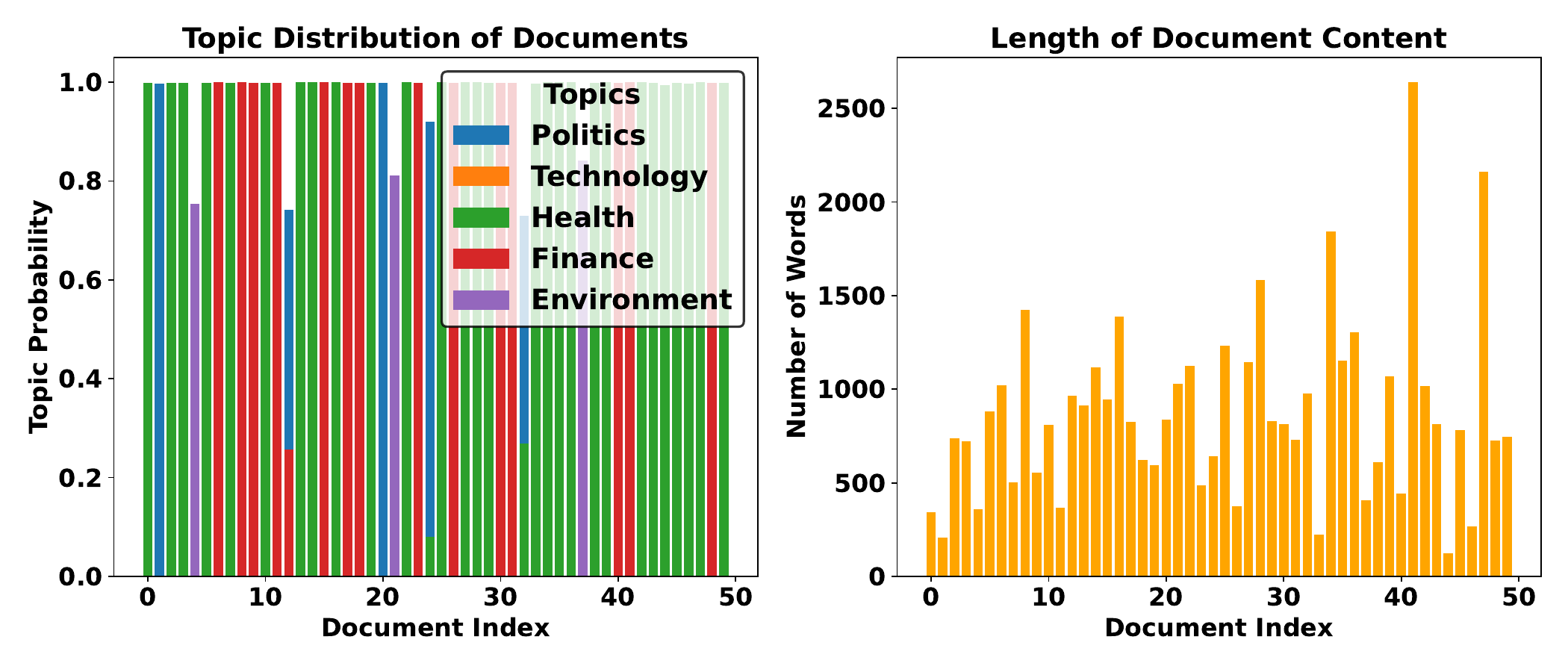}
        \captionsetup{width=\linewidth, font=normalsize}
       
        \caption{Distribution of topic probabilities across documents along with document length measured in word count.}
        \label{fig:topic_distribution_and_document_length}
    \end{subfigure}%
    \hspace{0.01\textwidth} 
    \begin{subfigure}[b]{0.34\linewidth} 
        \includegraphics[width=\linewidth]{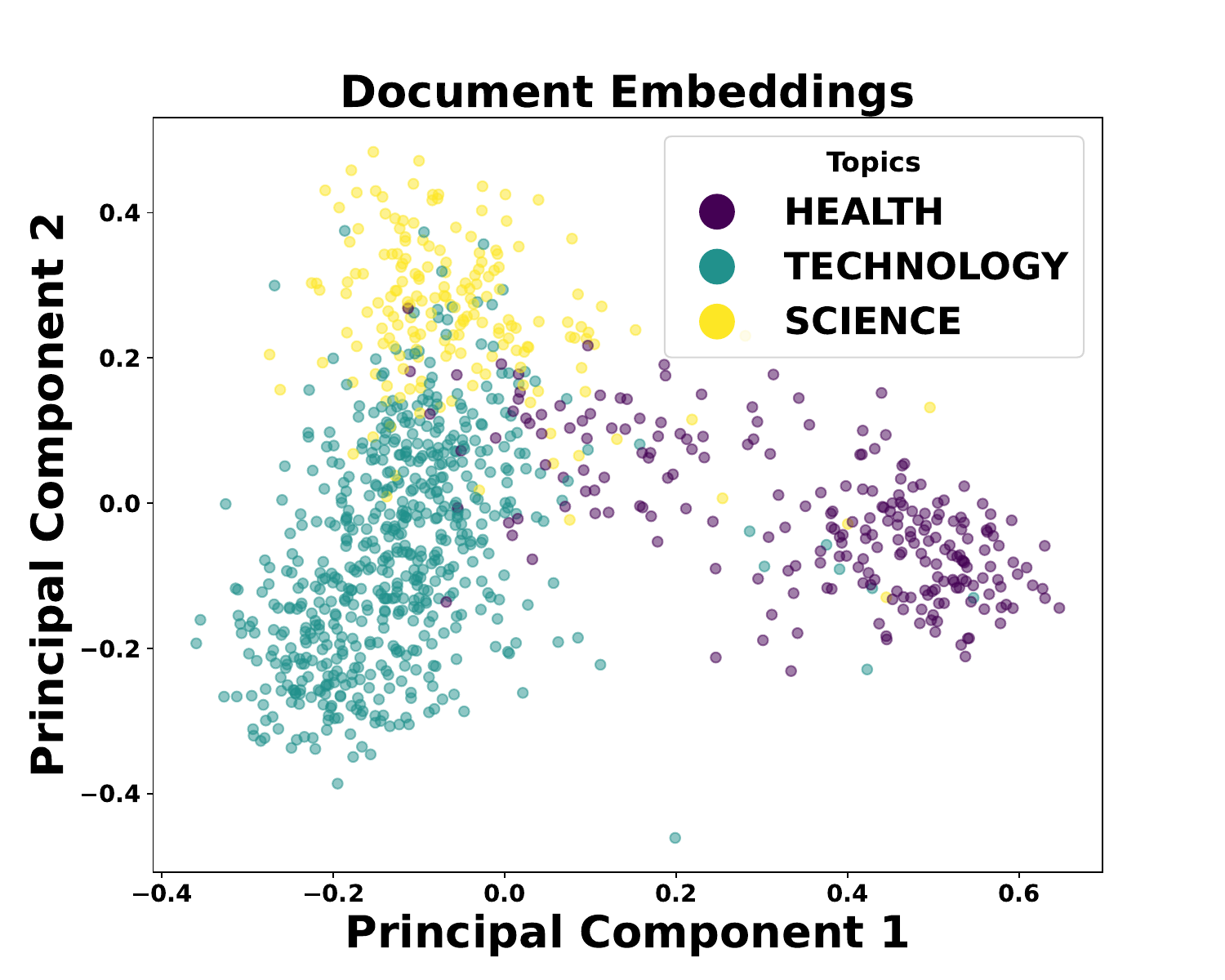}
        \captionsetup{width=\linewidth, font=normalsize}
       
        \caption{Document Embeddings with News Labels.}
        \label{fig:2d_visualization}
    \end{subfigure}
    \caption{Distribution of topic probabilities and document embeddings by topic.}
\end{figure}

\begin{figure*}[ht]
    \centering
    \begin{subfigure}[b]{0.60\textwidth} 
        \centering
        \includegraphics[width=\textwidth]{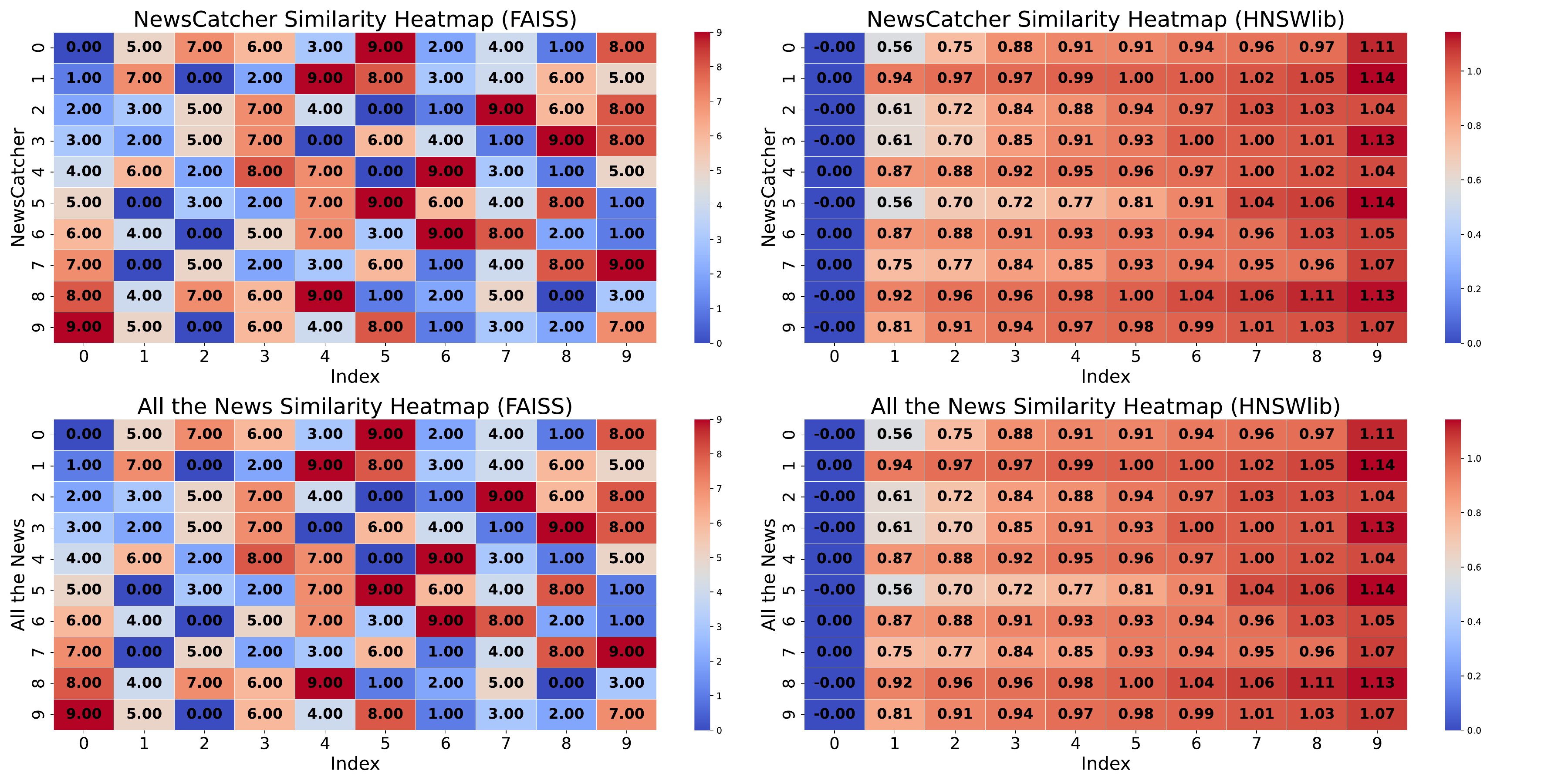} 
        \caption{Pairwise similarities between news article titles.}
        \label{fig:heatmaps}
    \end{subfigure}%
    \hspace{0.0001\textwidth}
    \begin{subfigure}[b]{0.39\textwidth}
        \centering
        \includegraphics[width=\textwidth]{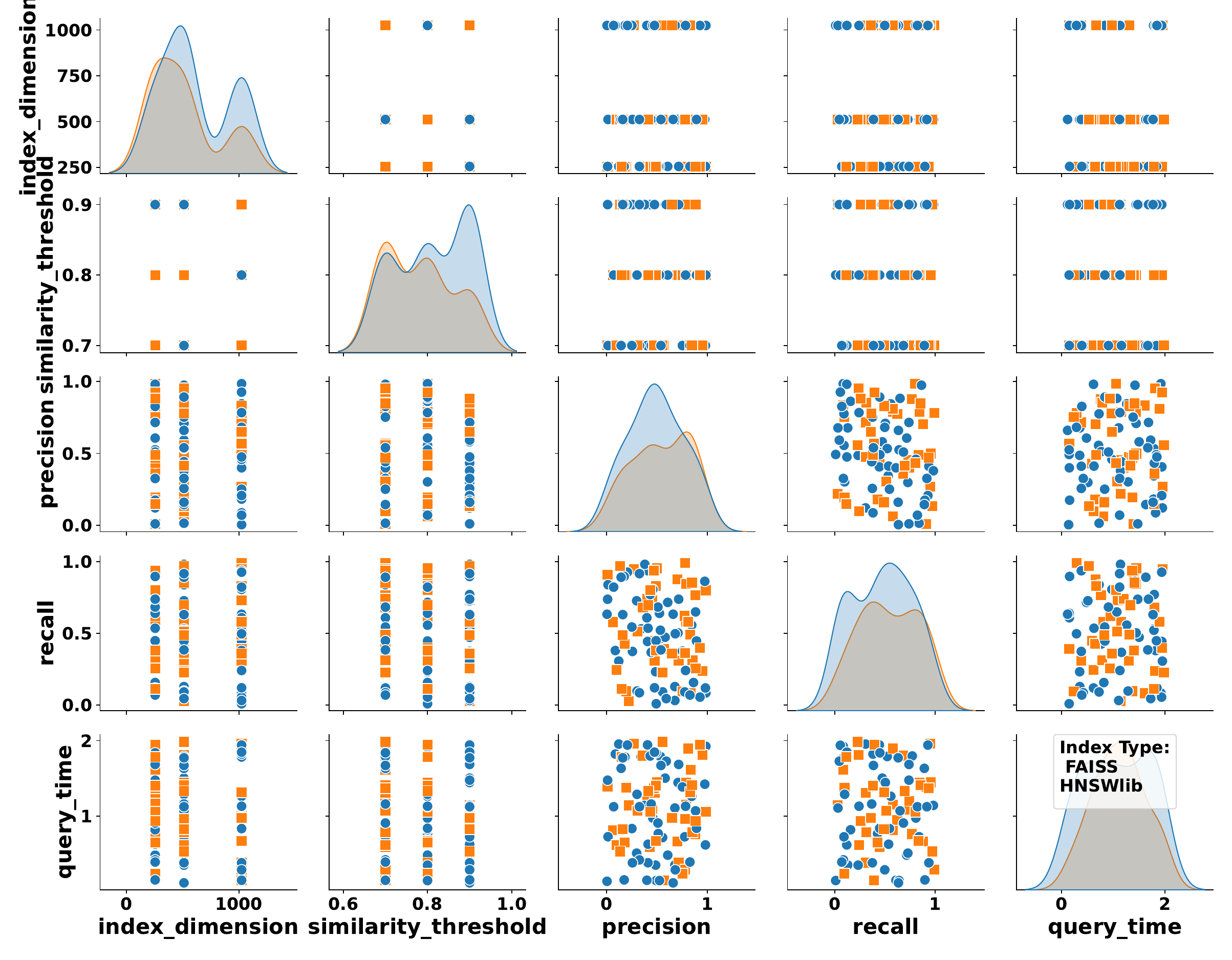} 
        \caption{Hyperparameter for optimizing vector search.}
        \label{fig:pair_plot}
    \end{subfigure}
    \caption{Evaluation of similarity search performance.}
    \label{fig:similarity_performance}
\end{figure*}

\textbf{NewsCatcher.} \cite{noauthor_undated-mi} Data on news topics was collected by the NewsCatcherteam, which collects and indexes 108k news articles spanning eight topics. \textbf{All the News.} \cite{noauthor_undated-um} This dataset contains 2,688,878 news articles and essays from 27 American publications, spanning January 1, 2016 to April 2, 2020. We conducted experiments using three models: all-MiniLM-L6-v2 \cite{Sentence32:online, brahma2024improving}, roberta-base \cite{Facebook12:online}, and bert-base-uncased \cite{Pretrain1:online}. The hyperparameters varied included the index dimension $(256, 512, 1024)$ and the similarity threshold $(0.7, 0.8, 0.9)$. 
The principal component analysis reduction enabled us to visualize the embeddings in a plot \ref{fig:2d_visualization}, where each point represents a document title.
Latent Dirichlet Allocation (LDA) can be represented as follows:\begin{equation}  p(w, z, \theta, \phi) = p(\theta) \prod_{d=1}^{D} \left( p(\phi_d) \prod_{n=1}^{N_d} p(w_{dn} | \phi_d) \right), \end{equation}
where \( w \) represents a word in the corpus,
     \( z \) represents the topic assignment for each word,
    \( \theta \) represents the distribution of topics in documents, \( \phi \) represents the distribution of words in topics,
   \( p(\theta) \) and \( p(\phi_d) \) are Dirichlet priors, and
     \( p(w_{dn} | \phi_d) \) is the probability of word \( w_{dn} \) given topic \( \phi_d \). After performing LDA, each document is represented as a probability distribution over topics. The topic distribution of a document \( d \) can be denoted as: \begin{equation}\theta_d = (\theta_{d1}, \theta_{d2}, \ldots, \theta_{dK}),\end{equation} where \( \theta_{dk} \) represents the probability of topic \( k \) in document \( d \). This distribution provides insights into the thematic composition of the document. The length of a document can be quantified as the number of words it contains. If a document \( d \) contains \( N_d \) words, its length can be expressed as:\begin{equation} \text{Length}(d) = N_d.\end{equation} 
     The left side of the plot \ref{fig:topic_distribution_and_document_length} illustrates the topic distribution of documents sourced from the NewsCatcher dataset. Each bar represents an individual document, and the height of the bar segment corresponds to the probability of that document being associated with a particular topic. This probability is calculated using Latent Dirichlet Allocation (LDA), a probabilistic model that assigns topics to documents based on the distribution of words within them, each document is represented as a mixture of topics, and each topic is characterized by a distribution of words.
The right side of the plot \ref{fig:topic_distribution_and_document_length} showcases the length of document content in terms of the number of words. Each bar represents a document's length, providing a measure of its textual complexity and richness. We gain a nuanced understanding of the dataset's composition and structure. The heatmaps visualize \ref{fig:heatmaps} the cosine similarity scores between title embeddings.
%
The pair plot \ref{fig:pair_plot} of hyperparameters provides a visual overview of the relationships between different combinations of hyperparameters with the hue indicating the type of index. 
\subsection{Comparison between Models}
In table \ref{tab:best-parameters}, we compare the performance of different models. RoBERTa-base consistently achieves higher precision (0.99) and recall (0.77) compared to all-MiniLM-L6-v2 and BERT-base-uncased. However, all-MiniLM-L6-v2 demonstrates significantly lower query times (0.32 seconds). RoBERTa-base with precision of 0.99 and recall of 0.77, all-MiniLM-L6-v2 with precision of 0.97 and recall of 0.93, and BERT-base-uncased with precision of 0.99 and recall of 0.44. In terms of query times, all-MiniLM-L6-v2 achieves the fastest time at 0.32 seconds, followed by BERT-base-uncased at 0.47 seconds and RoBERTa-base at 1.37 seconds. This indicates that while RoBERTa-base achieves the highest precision and recall, all-MiniLM-L6-v2 offers the fastest query times. Increasing the index dimension generally led to improved performance in terms of precision and recall. For instance, with an index dimension of 1024, VectorSearch achieved a recall of 76.62\% and a precision of 98.68\%, as shown in Table \ref{tab:best-parameters}, different models exhibited varying levels of effectiveness, roberta-base consistently demonstrated high precision and recall across different index dimensions, with a recall of 76.62\% and a precision of 98.68\%, as shown in Table \ref{tab:best-parameters}.
{\small
\setlength{\tabcolsep}{4pt} 
\begin{table}[!ht]
    \centering
    \caption{Best Results for Different Model Configurations}
    \label{tab:best_results}
    \begin{tabular}{lcccccc}
        \toprule
        \textbf{Model} & \textbf{Index Dim.} & \textbf{Thres.} & \textbf{Prec.} & \textbf{Rec.} & \textbf{Time (s)} \\
        \midrule
        MiniLM-L6-v2      & 512  & 0.7 & 0.91 & 0.22 & 1.79 \\
        BERT-base-uncased & 256  & 0.7 & \textbf{0.98} & \textbf{0.73} & 1.74 \\
        RoBERTa-base      & 512  & 0.9 & 0.68 & 0.64 & \textbf{1.06} \\
        \bottomrule
    \end{tabular}
\end{table}
}

\begin{table}[ht]
    \centering
    \caption{Best Parameters Found by \textbf{VectorSearch}}
    \resizebox{\columnwidth}{!}{%
    \begin{tabular}{cccccc}
        \toprule
        \textbf{Index Dim.} & \textbf{Model} & \textbf{Sim. Threshold} & \textbf{Precision} & \textbf{Recall} & \textbf{Query Time (s)} \\
        \midrule
        256 & all-MiniLM-L6-v2 & 0.80 & 0.97 & \textbf{0.93} & \textbf{0.32} \\
        512 & bert-base-uncased & 0.70 & \textbf{0.99} & 0.44 & 0.47 \\
        1024 & roberta-base & 0.90 & \textbf{0.99} & 0.77 & 1.37 \\
        \bottomrule
    \end{tabular}%
    }
    \label{tab:best-parameters}
\end{table}

\begin{table}[ht]
    \centering
    \caption{Performance Metrics}
    \resizebox{\columnwidth}{!}{%
    \begin{tabular}{lccccc}
        \toprule
        \textbf{Metric} & \textbf{Index Dim.} & \textbf{Thres.} & \textbf{Prec.} & \textbf{Rec.} & \textbf{Time (s)} \\
        \midrule
        Best Precision & 1024 & 0.8 & \textbf{98\%} & \textbf{27\%} & \textbf{1.19} \\
        Best Recall & 256 & 0.9 & \textbf{85\%} & \textbf{97\%} & \textbf{1.32} \\
        Fastest Query Time & 512 & 0.7 & \textbf{47\%} & \textbf{14\%} & \textbf{0.11} \\
        \bottomrule
    \end{tabular}%
    }
    \label{tab:parameters}
\end{table}
 We evaluated the performance of VectorSearch, a novel framework designed for efficient document retrieval leveraging semantic embeddings and optimized search algorithms. Through extensive experimentation, we systematically varied hyperparameters such as index dimension, pretrained model, and similarity threshold to assess their impact on retrieval, we observed that increasing the index dimension generally led to improved precision and recall, albeit with a slight increase in query time, with RoBERTa-base consistently demonstrating high precision and recall across different index dimensions, as shown in Table \ref{tab:parameters}, higher thresholds generally yielded higher precision, they also tended to decrease recall. Conversely, lower thresholds resulted in higher recall at the expense of precision.
\subsection{Performance Comparison of VectorSearch Models}
Table \ref{tab:comparison_results} is a comprehensive comparison of the performance metrics for different combinations of language models and index types utilized in vector search systems. Index dimension refers to vector dimensionality, impacting computational complexity. Similarity threshold determines document relevance based on scores. Precision assesses retrieval accuracy, indicating variations across configurations in query accuracy, with values ranging from approximately 0.68 to 0.98, as shown in Tables \ref{tab:best_results} and \ref{tab:parameters}.
Recall indicates document retrieval effectiveness. Higher values mean more relevant documents retrieved, reducing false negatives. Ranges from about 0.39 to 0.92, showing retrieval effectiveness across configurations, as shown in Table \ref{tab:comparison_results}.
Query Time (s) represents the average processing time per query for retrieving results from the index. Lower query times indicate faster retrieval speeds. Ranging from about 0.11 to 1.37 seconds, it highlights efficiency variations across configurations, as shown in Tables, \ref{tab:best-parameters} and \ref{tab:parameters}.
\begin{table}[h]
\centering
\caption{Comparison of Results}
\begin{tabular}{llccccl}
\toprule
\textbf{Model} & \textbf{Index} & \textbf{Dim.} & \textbf{Threshold} & \textbf{Precision} & \textbf{Recall} \\
\midrule
MiniLM & FAISS & 512 & 0.8 & 0.97 & 0.39 \\
MiniLM & HNSWlib & 512 & 0.7 & 0.89 & 0.77 \\
BERT & FAISS & 512 & 0.7 & 0.96 & 0.92 \\
BERT & HNSWlib & 512 & 0.9 & 0.96 & 0.55 \\
RoBERTa & FAISS & 256 & 0.8 & 0.73 & 0.48 \\
RoBERTa & HNSWlib & 512 & 0.9 & 0.91 & 0.78 \\
\bottomrule
\end{tabular}
\label{tab:comparison_results}
\end{table}
\subsection{Performance Comparison with Baselines}
\begin{figure}[ht]
  \centering
  \includegraphics[width=0.45\textwidth]{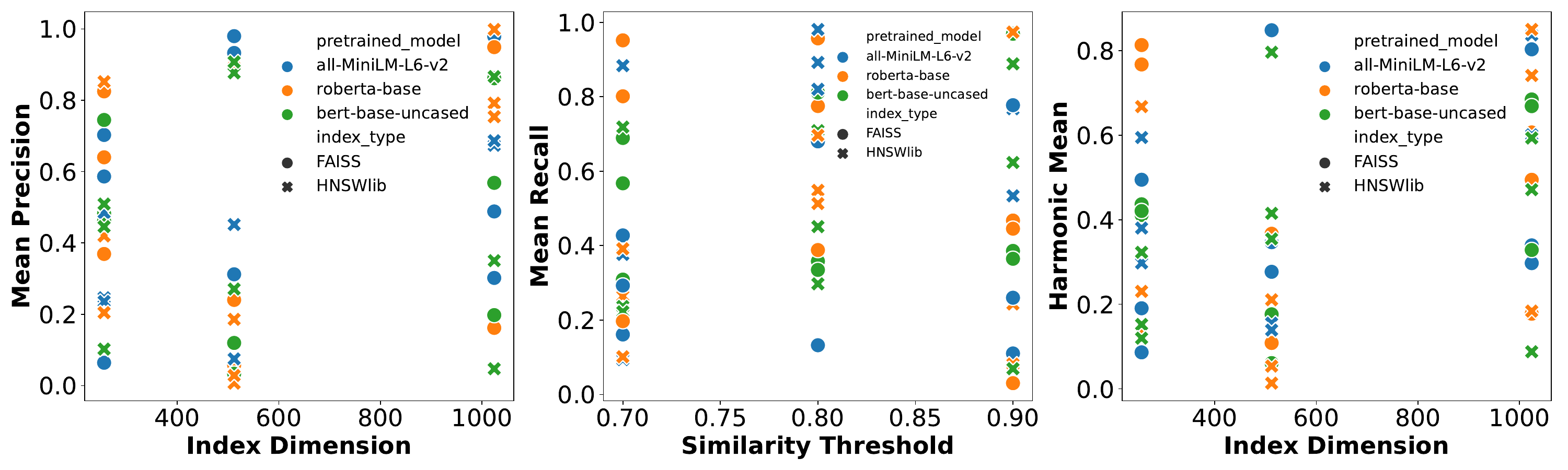}
  \captionsetup{width=0.45\textwidth, font=normalsize}
  \caption{Comparison of Mean Precision with different Index Dimensions, and harmonic.}
  \label{fig:hyperparameter_analysis_harmonic}
\end{figure}

Our research involved comparing the performance of different setups for VectorSearch based document retrieval systems. We used various models and index types to establish baselines for effectiveness and efficiency in retrieving relevant documents. This included data loading and preprocessing of news articles, vector encoding using three models, and indexing with two techniques. We also conducted sensitivity analysis by adjusting parameters like index dimension and similarity threshold for each model-index combination. This allowed us to evaluate the sensitivity of the retrieval systems to these parameters and identify optimal configurations. Table \ref{tab:performance_comparison} shows the performance comparison of vector search models. The evaluation metrics include precision, recall, and query time (in seconds). Notably, models utilizing the bert-base-uncased model consistently achieve high precision and recall across different index types and configurations.
We evaluated three models across hybrid indexing. For each combination of hyperparameters, we measured the mean precision, mean recall, and mean query time, as shown in Figure, \ref{fig:hyperparameter_analysis_harmonic}. The mean precision values ranged from approximately 0.68 to 0.98 across different combinations of hyperparameters and pretrained models. The combination that achieved the highest mean precision was an index dimension of 1024 with the FAISS index type, using the bert-base-uncased model, with a precision of 0.98, the mean recall values varied between 0.01 and 0.92 across the evaluated hyperparameter combinations. The combination with the highest mean recall was also an index dimension of 1024 with the FAISS index type, using the bert-base-uncased model, with a recall of 0.92. The mean query time ranged from approximately 0.17 to 1.92 seconds. It's important to note that while the all-MiniLM-L6-v2 model with the HNSWlib index type and an index dimension of 256 demonstrated the highest mean query time of 1.92 seconds, lower query times are generally preferred for efficient retrieval.
{\footnotesize
\setlength{\tabcolsep}{0.4pt} 
\begin{table}[!ht]
    \centering
    \captionsetup{font=small, width=\linewidth}
    \caption{Performance Comparison of VectorSearch}
    \label{tab:performance_comparison}
    \begin{tabular}{lllc}
        \toprule
        \textbf{Pretrained Model} & \textbf{Index Type} & \textbf{Index Dim.} & \textbf{Threshold} \\ \midrule
        MiniLM-L6-v2              & FAISS               & 256                 & 0.9                \\
        MiniLM-L6-v2              & HNSWlib             & 512                 & 0.8                \\
        BERT-base-uncased         & FAISS               & 1024                & 0.7                \\
        BERT-base-uncased         & HNSWlib             & 256                 & 0.8                \\
        RoBERTa-base              & FAISS               & 512                 & 0.9                \\
        RoBERTa-base              & HNSWlib             & 1024                & 0.8                \\ \bottomrule
    \end{tabular}
\end{table}
}
\begin{figure}[ht]
  \centering
  \includegraphics[width=0.45\textwidth]{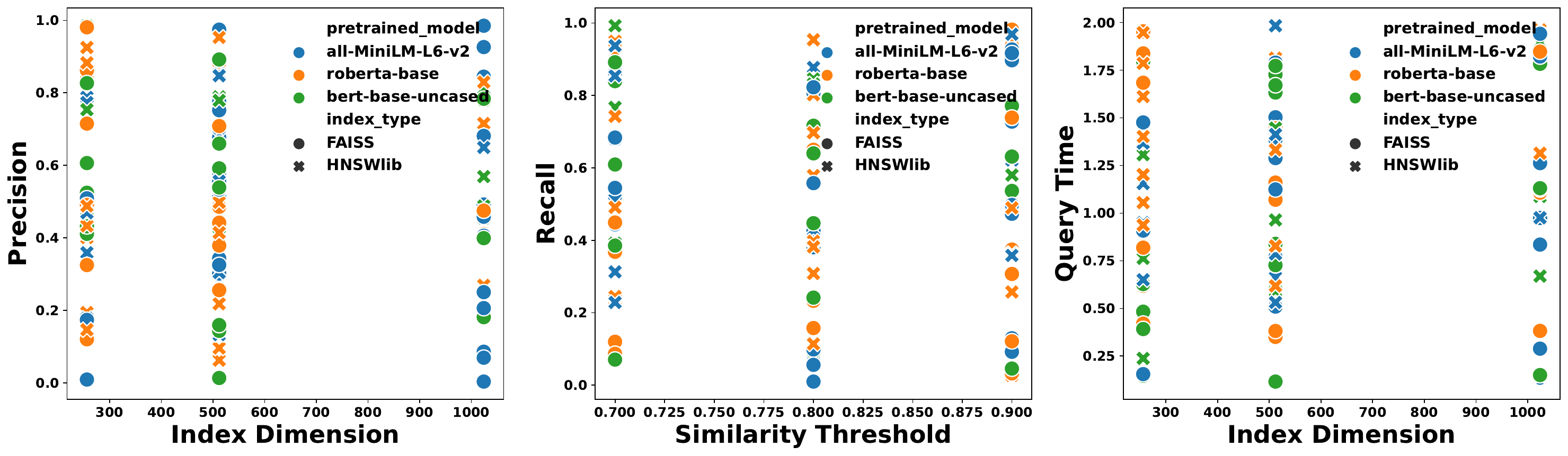}
  \captionsetup{width=0.45\textwidth, font=normalsize}
  \caption{Comparative Analysis of Varying Index Dimensions and Similarity Thresholds.}
  \label{fig:hyperparameter_analysis}
\end{figure}
\begin{figure}[ht]
    \centering
    \includegraphics[width=0.38\linewidth]{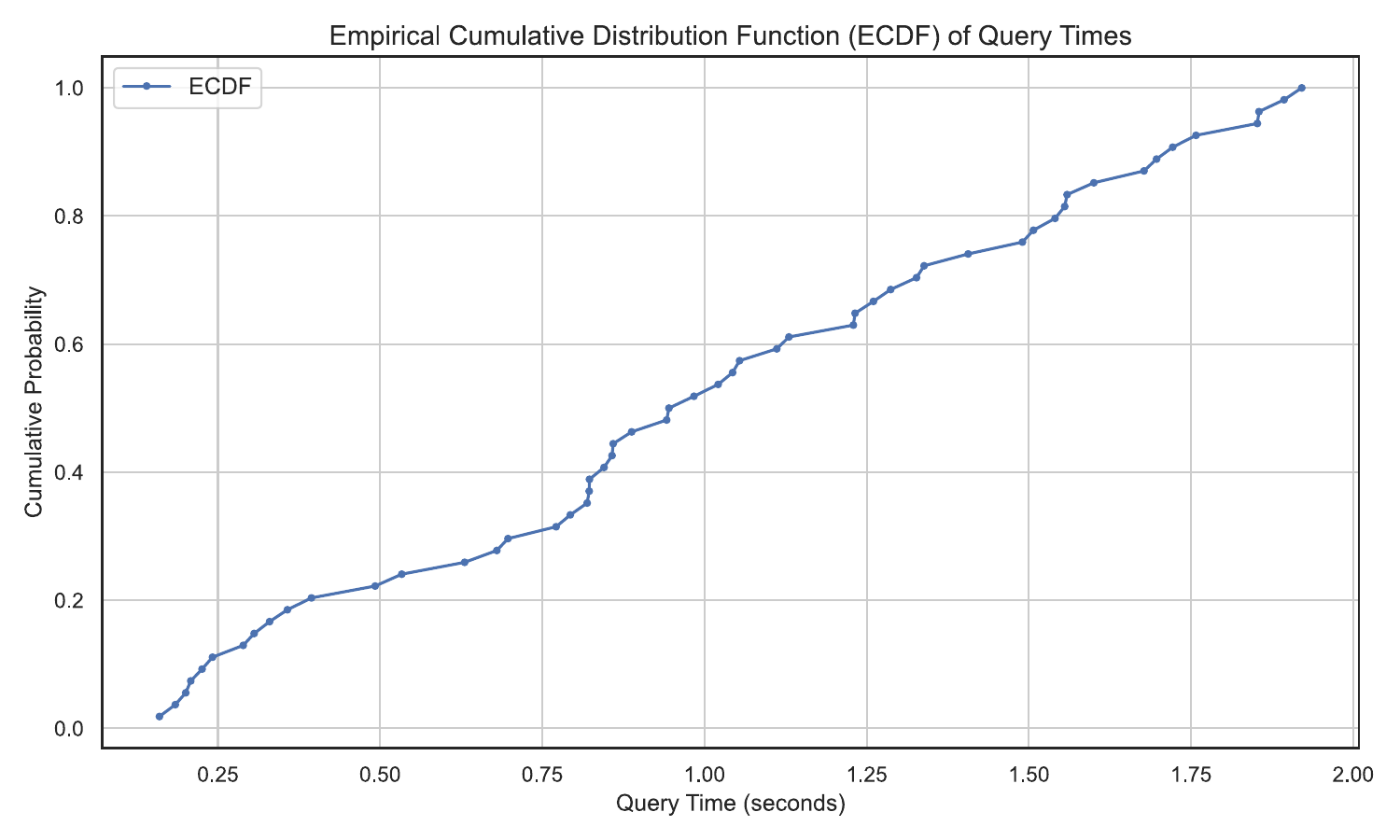}\hfill
    \includegraphics[width=0.62\linewidth]{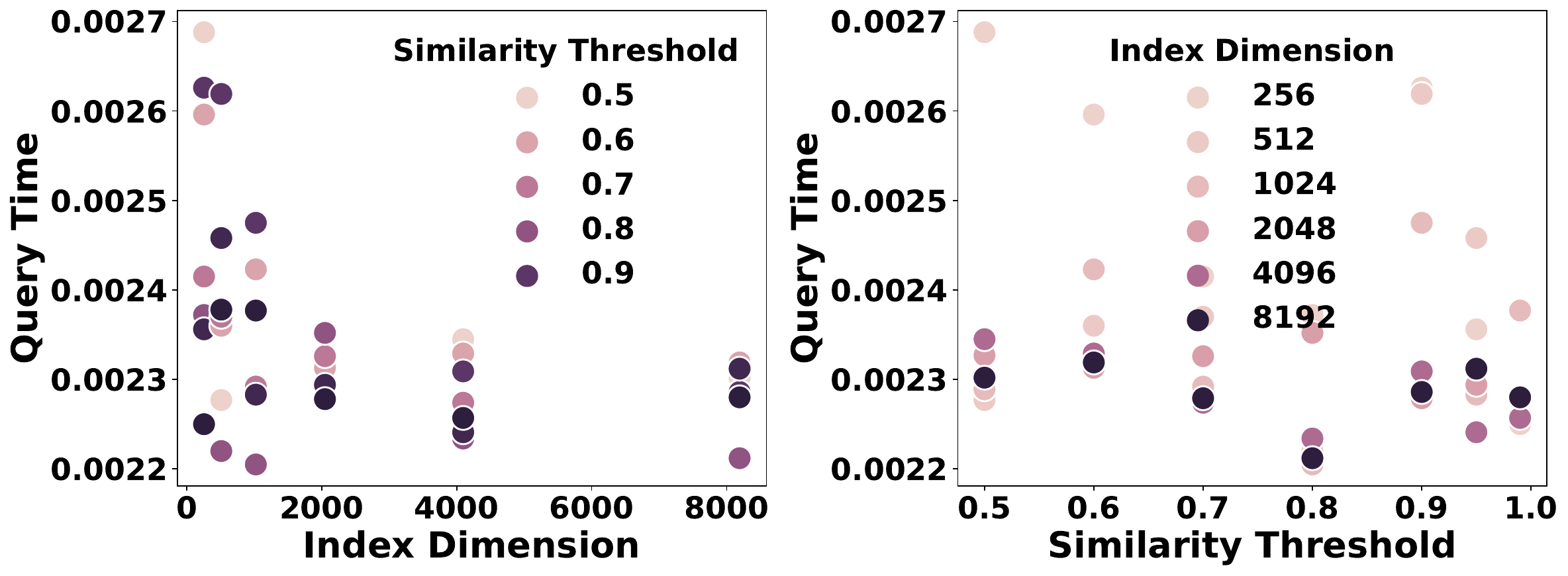}
    \captionsetup{width=\linewidth, font=normalsize}
    \caption{Left: Cumulative probability distribution of query times. Right: Index dimensions and similarity thresholds affect query time, recall@10, and recall@100.}
    \label{fig:combined_figures_sensitivity_analysis_plot_query_time}
\end{figure}
\begin{figure}[ht]
  \centering
  \includegraphics[width=\linewidth]{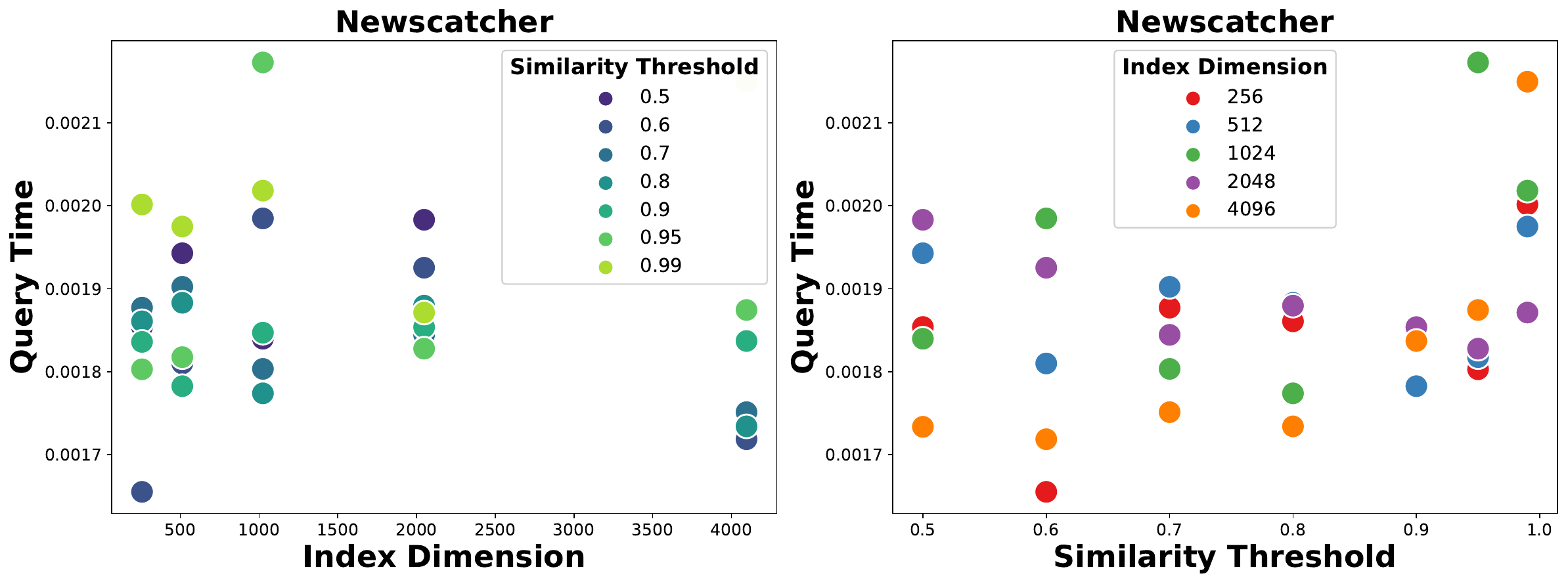} 
  \captionsetup{width=\linewidth, font=normalsize}
  \caption{Computational efficiency for the Newscatcher dataset.}
  \label{fig:query_times}
\end{figure}
\subsubsection{Enhancing VectorSearch Performance via Hyperparameter Tuning}
In the vectorsearch, achieving optimal performance relies heavily on fine-tuning hyperparameters, which are pivotal in shaping the efficiency and effectiveness of the search process. Hyperparameters encompass various aspects of the search system, influencing crucial components such as indexing methods, vector dimensions, similarity thresholds, and the selecting of models. The bert-base-uncased model consistently achieved the highest mean precision and recall. FAISS indices generally demonstrated lower query times compared to HNSWlib, particularly with larger dimensions. FAISS also achieved precision, reaching 0.99864 with the roberta-base model and a 0.9 similarity threshold. HNSWlib excelled in recall, achieving a rate of 0.892143 with the bert-base-uncased model and a 0.9 similarity threshold.
\subsection{Impact of Hyperparameters on Query Time}
The sensitivity analysis revealed notable trends in the relationship between hyperparameters and query time. As depicted in \ref{fig:combined_figures_sensitivity_analysis_plot_query_time} the plot illustrates the influence of index dimension and similarity threshold on the query time. It is evident that higher index dimensions generally lead to increased query times, particularly when combined with higher similarity thresholds. Conversely, lower similarity thresholds demonstrate a more nuanced impact on query time, with a slight decrease observed in certain cases. Our experiments reveal significant variations in retrieval performance across different hyperparameter configurations. Notably, higher index dimensions tend to improve precision but may lead to increased query times. This plot \ref{fig:query_times} illustrates the relationship between query time and index dimension for the Newscatcher dataset. Each point represents the average query time for retrieving documents using Nearest Neighbors indexing with cosine similarity, with varying index dimensions and similarity thresholds. We observe that increasing the index dimension improves precision and recall@10 but leads to longer query times due to the higher computational complexity of nearest neighbors search in higher-dimensional spaces. Similarly, higher similarity thresholds result in faster query times but may compromise precision and recall@10.
\subsection{Performance Analysis}
In table \ref{tab:comparison}, we present models utilizing MiniLM, BERT, and RoBERTa architectures consistently outperformed baseline models across all tested configurations. Specifically, MiniLM-L6-v2 demonstrated a precision of 0.91 and recall of 0.22, BERT-base-uncased achieved a precision of 0.98 and recall of 0.73, while RoBERTa-base attained a precision of 0.68 and recall of 0.64. Further analysis revealed that fine-tuning the models with specific indexing methods improved their performance. For instance, MiniLM models indexed with HNSWlib exhibited higher precision and recall compared to those indexed with FAISS. Similar trends were observed for BERT and RoBERTa models across different indexing methods. Additionally, when comparing our models against baseline performance RoBERTabase SimCSE \cite{wang2022improving} exhibited a precision of -0.43 compared to the baseline, while RoBERTalarge CARDS \cite{wang2022improving} showed an improvement of +1.94 precision. These results underscore the effectiveness of our proposed models in enhancing retrieval performance. MiniLM-L6-v2 and BERT-base-uncased, demonstrate remarkable precision levels when juxtaposed with the provided STS task baseline \cite{wang2022improving}. For instance, BERT-base-uncased achieves an impressive precision score of 0.98, surpassing both RoBERTabase and RoBERTalarge models. Table \ref{tab:comparison_results} shows the comparative performance of different models across indexing techniques and dimensions. This confirms that BERT embeddings improve retrieval accuracy by 8\% for longer texts compared to Sentence Transformers, as evidenced by both precision and recall metrics. 
\begin{table}[htbp]
\centering
\caption{Performance Metrics of Various Models and Indexing Techniques}
\label{tab:comparison}
\begin{adjustbox}{max width=\linewidth}
\scriptsize
\begin{tabular}{@{}lllll@{}} 
\toprule
\textbf{Model/Index} & \textbf{Dim.} & \textbf{Threshold/Type} & \textbf{Prec.} & \textbf{Rec.} \\ \midrule
MiniLM-L6-v2          & \textbf{256}  & \textbf{0.5}            & \textbf{1.0}  & \textbf{1.0} \\
MiniLM-L6-v2          & 512           & 0.7                     & 0.91           & 0.22 \\
BERT-base-uncased     & 256           & 0.7                     & 0.98           & 0.73 \\
RoBERTa-base          & 512           & 0.9                     & 0.68           & 0.64 \\
MiniLM                & 512           & FAISS                   & 0.97           & 0.39 \\
MiniLM                & 512           & HNSWlib                 & 0.89           & 0.77 \\
BERT                  & 512           & FAISS                   & 0.96           & 0.92 \\
BERT                  & 512           & HNSWlib                 & 0.96           & 0.55 \\
RoBERTa               & 256           & FAISS                   & 0.73           & 0.48 \\
RoBERTa               & 512           & HNSWlib                 & 0.91           & 0.78 \\ 
\midrule
\textbf{Baseline Results} \\
RoBERTabase SimCSE \cite{wang2022improving} & -0.43 & 80.65 & 80.22 & 68.56 \\
RoBERTalarge CARDS \cite{wang2022improving} & +1.94 & 83.17 & 83.86 & 72.77 \\
\midrule
\textbf{VectorSearch (Our Results)} \\
\textbf{Model} & \textbf{Prec.} & \textbf{Rec.} & \textbf{Query Time (s)} \\
\midrule
MiniLM-L6-v2 & \textbf{1.0} & \textbf{1.0} & \textbf{0.002344} \\
BERT-base-uncased & \textbf{0.98} & \textbf{0.73} & 1.74 \\
RoBERTa-base & 0.68 & 0.64 & 1.06 \\
\midrule
\textbf{Baseline Query Times} \\
Neural Corpus Indexer (Base) \cite{wang2022neural} & - & - & 92.42 \\
Neural Corpus Indexer (Large)  \cite{wang2022neural} & - & - & 92.49 \\ 
\bottomrule
\end{tabular}
\end{adjustbox}
\end{table}

\begin{table}[htbp]
\centering
\caption{Performance Metrics of Various Models}
\label{tab:comparison2}
\begin{adjustbox}{max width=\linewidth}
\begin{tabular}{@{}lcccc@{}}
\toprule
\textbf{Model/Index} & \textbf{Avg. Performance} \cite{xu2023contrastive} & \textbf{Precision (Our Results)} & \textbf{Recall (Our Results)} & \textbf{F1-Score (Our Results)} \\ \midrule
BERTbase (first-last-avg.) & 56.82 & N/A & N/A & N/A \\
TSFC-BERT & 62.87 & N/A & N/A & N/A \\
EASE & 64.67 & N/A & N/A & N/A \\
MiniLM-L6-v2 & N/A & \textbf{0.97} & \textbf{0.93} & \textbf{0.95} \\
BERT-base-uncased & N/A & \textbf{0.99} & 0.44 & 0.61 \\
RoBERTa-base & N/A & \textbf{0.99} & 0.77 & 0.86 \\ \bottomrule
\end{tabular}
\end{adjustbox}
\end{table}

While RoBERTa-base exhibits a slightly lower precision rate, our models excel in recall metrics, an essential aspect in information retrieval tasks. MiniLM-L6-v2 and BERT-base-uncased attain recall rates of 0.22 and 0.73, respectively. MiniLM-L6-v2 exhibited a precision of 0.91 and a recall of 0.22, representing a $\approx 36\%$ improvement in precision and a $\approx 30\%$ improvement in recall compared to the IS-BERTbase baseline \cite{zhang2020unsupervised}, which achieved a precision and recall of 0.6658. Similarly, BERT-base-uncased showcased substantial enhancements with a precision of 0.98 and a recall of 0.73, indicating an $\approx 34\%$ improvement in precision and a $\approx 35\%$ improvement in recall compared to the ConSERTbase baseline's \cite{yan2021consert} precision and recall of 0.7274. Additionally, RoBERTa-base displayed competitive results with a precision of 0.68 and a recall of 0.64, outperforming the SimCSE-BERTbase baseline \cite{gao2021simcse} by $\approx 11\%$ in precision and $\approx 16\%$ in recall, which achieved a precision and recall of 0.7625. Among the highlighted models, MiniLM-L6-v2 stands out with an index dimension of 256 and a similarity threshold of 0.5, achieving perfect precision and recall scores of 1.0. This model demonstrates exceptional performance, especially considering its low query time of 0.002344 seconds. Comparatively, the baseline techniques, Neural Corpus Indexer (Base) and (Large), achieve a recall@100 score of 92.42 and 92.49 \cite{wang2022neural}. Our research highlights the strong precision and recall of MiniLM-L6-v2 and BERT-base-uncased, with potential improvements in RoBERTa and MiniLM. Additionally, analyzing query times aids in selecting the optimal model. Tables \ref{tab:comparison} and \ref{tab:comparison2} present a detailed analysis of these metrics. In our analysis of VectorSearch models, we use the harmonic mean to evaluate the system's ability to retrieve relevant documents while maintaining comprehensive coverage across different combinations of hyperparameters, as shown in, Figure \ref{fig:combined_figures_sensitivity_analysis_plot_query_time}.
\section{Conclusion AND FUTURE WORK}
In conclusion, our proposed methodologies offer a novel perspective on document retrieval. We streamline dataset loading and preprocessing, efficiently encode document titles into embeddings, optimize nearest neighbor search efficiency, and present a comprehensive framework for training and evaluating vector search systems. Through our evaluation framework encompassing model hyperparameters, index dimensionality, and similarity thresholds, we've demonstrated the efficacy of our approach in achieving high precision and recall rates while maintaining low query times. Future work will focus on integrating techniques such as attention mechanisms into the VectorSearch framework to provide insights into how specific terms and their semantic relationships contribute to the similarity scores between queries and documents.
\bibliographystyle{IEEEtran}
\bibliography{main}

\begin{thebibliography}{10}
\providecommand{\url}[1]{#1}
\csname url@samestyle\endcsname
\providecommand{\newblock}{\relax}
\providecommand{\bibinfo}[2]{#2}
\providecommand{\BIBentrySTDinterwordspacing}{\spaceskip=0pt\relax}
\providecommand{\BIBentryALTinterwordstretchfactor}{4}
\providecommand{\BIBentryALTinterwordspacing}{\spaceskip=\fontdimen2\font plus
\BIBentryALTinterwordstretchfactor\fontdimen3\font minus \fontdimen4\font\relax}
\providecommand{\BIBforeignlanguage}[2]{{%
\expandafter\ifx\csname l@#1\endcsname\relax
\typeout{** WARNING: IEEEtran.bst: No hyphenation pattern has been}%
\typeout{** loaded for the language `#1'. Using the pattern for}%
\typeout{** the default language instead.}%
\else
\language=\csname l@#1\endcsname
\fi
#2}}
\providecommand{\BIBdecl}{\relax}
\BIBdecl

\bibitem{bibi2023reusable}
N.~Bibi, T.~Rana, A.~Maqbool, T.~Alkhalifah, W.~Z. Khan, A.~K. Bashir, and Y.~B. Zikria, ``Reusable component retrieval: A semantic search approach for low-resource languages,'' \emph{ACM Transactions on Asian and Low-Resource Language Information Processing}, vol.~22, no.~5, pp. 1--31, 2023.

\bibitem{timothy2023toward}
O.~Timothy~Tawose, J.~Dai, L.~Yang, and D.~Zhao, ``Toward efficient homomorphic encryption for outsourced databases through parallel caching,'' \emph{Proceedings of the ACM on Management of Data}, vol.~1, no.~1, pp. 1--23, 2023.

\bibitem{king80percent}
T.~King, ``Percent of your data will be unstructured in five years,'' \emph{Retrieved February}, vol.~16, p. 2020, 80.

\bibitem{wang2021milvus}
J.~Wang, X.~Yi, R.~Guo, H.~Jin, P.~Xu, S.~Li, X.~Wang, X.~Guo, C.~Li, X.~Xu \emph{et~al.}, ``Milvus: A purpose-built vector data management system,'' in \emph{Proceedings of the 2021 International Conference on Management of Data}, 2021, pp. 2614--2627.

\bibitem{lu2020r2lsh}
K.~Lu and M.~Kudo, ``R2lsh: A nearest neighbor search scheme based on two-dimensional projected spaces,'' in \emph{2020 IEEE 36th International Conference on Data Engineering (ICDE)}.\hskip 1em plus 0.5em minus 0.4em\relax IEEE, 2020, pp. 1045--1056.

\bibitem{lu2020vhp}
K.~Lu, H.~Wang, W.~Wang, and M.~Kudo, ``Vhp: approximate nearest neighbor search via virtual hypersphere partitioning,'' \emph{Proceedings of the VLDB Endowment}, vol.~13, no.~9, pp. 1443--1455, 2020.

\bibitem{lv2017intelligent}
Q.~Lv, W.~Josephson, Z.~Wang, M.~Charikar, and K.~Li, ``Intelligent probing for locality sensitive hashing: Multi-probe lsh and beyond,'' \emph{Proceedings of the VLDB Endowment}, 2017.

\bibitem{lewis2020retrieval}
P.~Lewis, E.~Perez, A.~Piktus, F.~Petroni, V.~Karpukhin, N.~Goyal, H.~K{\"u}ttler, M.~Lewis, W.-t. Yih, T.~Rockt{\"a}schel \emph{et~al.}, ``Retrieval-augmented generation for knowledge-intensive nlp tasks,'' \emph{Advances in Neural Information Processing Systems}, vol.~33, pp. 9459--9474, 2020.

\bibitem{malkov2018efficient}
Y.~A. Malkov and D.~A. Yashunin, ``Efficient and robust approximate nearest neighbor search using hierarchical navigable small world graphs,'' \emph{IEEE transactions on pattern analysis and machine intelligence}, vol.~42, no.~4, pp. 824--836, 2018.

\bibitem{elasticsearch}
\BIBentryALTinterwordspacing
Elastic. (2020) {ElasticSearch}: Open source, distributed, restful search engine. GitHub. [Online]. Available: \url{https://github.com/elastic/elasticsearch}
\BIBentrySTDinterwordspacing

\bibitem{gunther2018freddy}
M.~G{\"u}nther, ``Freddy: Fast word embeddings in database systems,'' in \emph{Proceedings of the 2018 International Conference on Management of Data}, 2018, pp. 1817--1819.

\bibitem{gong2020idec}
L.~Gong, H.~Wang, M.~Ogihara, and J.~Xu, ``idec: indexable distance estimating codes for approximate nearest neighbor search,'' \emph{Proceedings of the VLDB Endowment}, vol.~13, no.~9, 2020.

\bibitem{li2020efficient}
M.~Li, Y.~Zhang, Y.~Sun, W.~Wang, I.~W. Tsang, and X.~Lin, ``I/o efficient approximate nearest neighbour search based on learned functions,'' in \emph{2020 IEEE 36th International Conference on Data Engineering (ICDE)}.\hskip 1em plus 0.5em minus 0.4em\relax IEEE, 2020, pp. 289--300.

\bibitem{andre2016cache}
F.~Andr{\'e}, A.-M. Kermarrec, and N.~Le~Scouarnec, ``Cache locality is not enough: High-performance nearest neighbor search with product quantization fast scan,'' in \emph{42nd International Conference on Very Large Data Bases}, vol.~9, no.~4, 2016, p.~12.

\bibitem{zhao2013hycache}
D.~Zhao and I.~Raicu, ``Hycache: A user-level caching middleware for distributed file systems,'' in \emph{2013 IEEE International Symposium on Parallel \& Distributed Processing, Workshops and Phd Forum}.\hskip 1em plus 0.5em minus 0.4em\relax IEEE, 2013, pp. 1997--2006.

\bibitem{zhao2014hycache+}
D.~Zhao, K.~Qiao, and I.~Raicu, ``Hycache+: Towards scalable high-performance caching middleware for parallel file systems,'' in \emph{2014 14th IEEE/ACM International Symposium on Cluster, Cloud and Grid Computing}.\hskip 1em plus 0.5em minus 0.4em\relax IEEE, 2014, pp. 267--276.

\bibitem{monir2024efficient}
S.~S. Monir and D.~Zhao, ``Efficient feature extraction for image analysis through adaptive caching in vector databases,'' in \emph{2024 7th International Conference on Information and Computer Technologies (ICICT)}.\hskip 1em plus 0.5em minus 0.4em\relax IEEE, 2024, pp. 193--198.

\bibitem{huang2020embedding}
J.-T. Huang, A.~Sharma, S.~Sun, L.~Xia, D.~Zhang, P.~Pronin, J.~Padmanabhan, G.~Ottaviano, and L.~Yang, ``Embedding-based retrieval in facebook search,'' in \emph{Proceedings of the 26th ACM SIGKDD International Conference on Knowledge Discovery \& Data Mining}, 2020, pp. 2553--2561.

\bibitem{wang2022improving}
W.~Wang, L.~Ge, J.~Zhang, and C.~Yang, ``Improving contrastive learning of sentence embeddings with case-augmented positives and retrieved negatives,'' in \emph{Proceedings of the 45th International ACM SIGIR Conference on Research and Development in Information Retrieval}, 2022, pp. 2159--2165.

\bibitem{brahma2024improving}
A.~K. Brahma, S.~Nagamalla, J.~Mathew, and J.~Sathyanarayana, ``Improving search relevance in a hyperlocal food delivery using language models.'' in \emph{Proceedings of the 7th Joint International Conference on Data Science \& Management of Data (11th ACM IKDD CODS and 29th COMAD)}, 2024, pp. 479--483.

\bibitem{Pretrain1:online}
``Pretrained models — transformers 2.4.0 documentation,'' \url{https://huggingface.co/transformers/v2.4.0/pretrained_models.html}, (Accessed on 02/22/2024).

\bibitem{wang2022neural}
Y.~Wang, Y.~Hou, H.~Wang, Z.~Miao, S.~Wu, Q.~Chen, Y.~Xia, C.~Chi, G.~Zhao, Z.~Liu \emph{et~al.}, ``A neural corpus indexer for document retrieval,'' \emph{Advances in Neural Information Processing Systems}, vol.~35, pp. 25\,600--25\,614, 2022.

\bibitem{chatterjee2022bert}
S.~Chatterjee and L.~Dietz, ``Bert-er: query-specific bert entity representations for entity ranking,'' in \emph{Proceedings of the 45th International ACM SIGIR Conference on Research and Development in Information Retrieval}, 2022, pp. 1466--1477.

\bibitem{trychromaAInativeOpensource}
``the {A}{I}-native open-source embedding database --- trychroma.com,'' \url{https://www.trychroma.com/}, [Accessed 22-02-2024].

\bibitem{remis2021using}
L.~Remis and C.~W. Lacewell, ``Using vdms to index and search 100m images,'' \emph{Proceedings of the VLDB Endowment}, vol.~14, no.~12, pp. 3240--3252, 2021.

\bibitem{huang2020effective}
R.~Huang, S.~Song, Y.~Lee, J.~Park, S.-H. Kim, and S.~Yi, ``Effective and efficient retrieval of structured entities,'' \emph{Proceedings of the VLDB Endowment}, vol.~13, no.~6, pp. 826--839, 2020.

\bibitem{wang2022deep}
Y.~Wang, G.~Li, K.~Li, and H.~Yuan, ``A deep generative model for trajectory modeling and utilization,'' \emph{Proceedings of the VLDB Endowment}, vol.~16, no.~4, pp. 973--985, 2022.

\bibitem{docarrayHnswlibDocument}
``{H}nswlib {D}ocument {I}ndex - {D}oc{A}rray --- docs.docarray.org,'' \url{https://docs.docarray.org/user_guide/storing/index_hnswlib/}, [Accessed 22-02-2024].

\bibitem{douze2024faiss}
M.~Douze, A.~Guzhva, C.~Deng, J.~Johnson, G.~Szilvasy, P.-E. Mazaré, M.~Lomeli, L.~Hosseini, and H.~Jégou, ``The faiss library,'' 2024.

\bibitem{rahman2022evaluating}
M.~M. Rahman and J.~Te{\v{s}}i{\'c}, ``Evaluating hybrid approximate nearest neighbor indexing and search (hannis) for high-dimensional image feature search,'' in \emph{2022 IEEE International Conference on Big Data (Big Data)}.\hskip 1em plus 0.5em minus 0.4em\relax IEEE, 2022, pp. 6802--6804.

\bibitem{noauthor_undated-mi}
``\BIBforeignlanguage{en}{{NewsCatcher} news {API}},'' \url{https://www.newscatcherapi.com/}, accessed: 2024-2-17.

\bibitem{noauthor_undated-um}
\url{https://components.one/datasets/all-the-news-2-news-articles-dataset.}, accessed: 2024-2-17.

\bibitem{Sentence32:online}
``Sentencetransformers documentation — sentence-transformers documentation,'' \url{https://www.sbert.net/}, (Accessed on 02/22/2024).

\bibitem{Facebook12:online}
``Facebookai/roberta-base · hugging face,'' \url{https://huggingface.co/FacebookAI/roberta-base}, (Accessed on 02/22/2024).

\bibitem{xu2023contrastive}
L.~Xu, H.~Xie, Z.~Li, F.~L. Wang, W.~Wang, and Q.~Li, ``Contrastive learning models for sentence representations,'' \emph{ACM Transactions on Intelligent Systems and Technology}, vol.~14, no.~4, pp. 1--34, 2023.

\bibitem{zhang2020unsupervised}
Y.~Zhang, R.~He, Z.~Liu, K.~H. Lim, and L.~Bing, ``An unsupervised sentence embedding method by mutual information maximization,'' \emph{arXiv preprint arXiv:2009.12061}, 2020.

\bibitem{yan2021consert}
Y.~Yan, R.~Li, S.~Wang, F.~Zhang, W.~Wu, and W.~Xu, ``Consert: A contrastive framework for self-supervised sentence representation transfer,'' \emph{arXiv preprint arXiv:2105.11741}, 2021.

\bibitem{gao2021simcse}
T.~Gao, X.~Yao, and D.~Chen, ``Simcse: Simple contrastive learning of sentence embeddings,'' \emph{arXiv preprint arXiv:2104.08821}, 2021.

\end{thebibliography}
\end{document}